\documentclass[10pt,letterpaper]{article}
\usepackage[latin1]{inputenc}
\usepackage{amsmath}
\usepackage{amsfonts}
\usepackage{amssymb}
\usepackage{geometry}
\usepackage{graphicx}
\usepackage{bm}
\usepackage[pdftex,pdfpagemode=UseNone,pdfstartpage=1,pdfstartview=FitH,
	pdfauthor={Kane},
	pdftitle={Levitated Spinning Graphene}]{hyperref}
\usepackage{color}

\def\persec{$\mathrm{sec^{-1}}$}
\def\permin{$\mathrm{min^{-1}}$}
\def\perm{$\mathrm{m^{-1}}$}
\def\perkg{$\mathrm{kg^{-1}}$}
\def\perg{$\mathrm{g^{-1}}$}
\def\perL{$\mathrm{L^{-1}}$}
\def\peramu{$\mathrm{amu^{-1}}$}
\def\2D{\mathrm{2D}}
\def\3D{\mathrm{3D}}
\def\kB{k_\mathrm{B}}
\def\gv{\gamma_v}
\def\gw{\gamma_\omega}
\def\Wtrap{\Omega_\mathrm{t}}
\def\wL{\omega_\mathrm{L}}
\def\wg{\omega_\mathrm{g}}
\def\Vtip{V_\mathrm{tip}}
\def\Vin{V_\mathrm{in}}
\def\Vout{V_\mathrm{out}}
\def\dcs{\left(\frac{d \sigma}{d \Omega}\right)}
\def\N2{N$_2$}

\begin{document}

\hsize\textwidth\columnwidth\hsize\csname@twocolumnfalse\endcsname

\title{Levitated Spinning Graphene}

\author{B. E. Kane\thanks{e-mail: bekane@umd.edu} \\
Laboratory for Physical Sciences and Joint Quantum Institute\\
University of Maryland\\College Park, MD 20742 USA}

\date{\today}

\maketitle

\begin{abstract}
A method is described for levitating micron-sized few layer graphene flakes in
a quadrupole ion trap.  Starting from a liquid suspension containing graphene,
charged flakes are injected into the trap using the electrospray ionization
technique and are probed optically.  At micro-torr pressures, torques from circularly polarized 
light cause the levitated particles to rotate at frequencies $>$1\,MHz, which can be inferred from
modulation of light scattering off the rotating flake when an electric field resonant
with the rotation rate is applied.   Possible applications of 
these techniques will be presented, both to fundamental measurements of the mechanical
and electronic properties of graphene and to
new approaches to graphene crystal growth, modification and manipulation.

\end{abstract}

\medskip
\medskip

\section{Introduction}
Since its discovery in 2004 \cite{Novoselov2004}, graphene has received a tremendous amount of attention directed both towards
understanding its fundamental properties and seeking applications for this new material \cite{Geim2009}\cite{Das_Sarma2010}\cite{Soldano2010}.
From the physics perspective,
graphene is the first truly two dimensional (2D) system, with electronic, mechanical, and
thermodynamic properties all determined by the structure of a single sheet of carbon atoms.  It seems
likely that the revolutionary applications for graphene in the future will make use in
some way of the unique properties of an intrinsically 2D system.

While graphene is an ideal 2D material, it nonetheless must be coupled to the outside world
in real experiments and, in practise, attached to a 3D substrate in
some way.  By now it is known that the substrate can limit mobility of graphene electrons 
\cite{Chen2008}\cite{Neugebauer2009}\cite{Dean2010},
and consequently the recent experiments demonstrating the fractional quantum Hall effect
in graphene \cite{Bolotin2009}\cite{Du2009}
were conducted on samples that were suspended (i.e. the substrate was etched away) in the region of measurement.
Locally suspended samples were also used for the first measurements of the mechanical properties of 
graphene \cite{Bunch2007}\cite{Lee2008}, but it is probable that loss mechanisms and
residual stresses in these experiments are determined by the substrate \cite{Garcia-Sanchez2008}.  The thermodynamic measurements
of graphene are perhaps most likely to be hindered by the presence of a substrate, since
the expected melting temperature of graphene is in excess of 3000\,K, higher than that of any
other material.  

It is possible to avoid coupling to a substrate altogether if graphene is levitated using
the particle trapping technologies that have been perfected in recent decades.  Indeed, optical
trapping of graphene suspended in solution has recently been demonstrated \cite{Marago2010}.
Diamagnetic trapping is possible for graphite \cite{Simon2000}.  For measurements of graphene, however,
quadrupole (ion) trapping has several advantages:  it is compatible with
ultrahigh vacuum (UHV), low, and high temperature measurements; it can provide tight confinement of very
small graphene particles; finally, ion trapping has the advantage
that charged graphene flakes will tend to remain flat when levitated due to
electrostatic repulsion. Interestingly, ion trap techniques have been applied previously to
\emph{graphite} particles for studies of interstellar dust \cite{Krauss2004}\cite{Spann2001}\cite{Abbas2004}\cite{Abbas2006}.  
Ion trapping of micron-sized particles has also been developed as a test-bed for
quantum information processing techniques \cite{Pearson2006}\cite{Clark2009} and for materials studies of
small particles \cite{Cai2002}\cite{Schlemmer2004}.

Below, I describe a quadrupole trap optimized for graphene measurements and present
preliminary data.  Graphene flakes originally suspended in liquid are injected into the trap vacuum chamber
using the electrospray ionization technique \cite{Fenn2003}.
Trapped flakes are detected optically.  The optical absorption
and low mass of the flakes mean that the particles absorb angular momentum from circularly polarized light
\cite{Friese1996}\cite{Xu2008}
and begin to spin rapidly at low pressure ($p \cong$1\,$\mu$torr).  This spinning is directly demonstrated by
the observation of rotational resonance - the modulation of optical scattering from the
particle at a well defined frequency of an applied electric field - at frequencies above 1\,MHz.  This
high rotation frequency, facilitated by the ability of graphene to withstand centrifugal tension
during rotation, is, to the author's knowledge, the largest ever measured for a macroscopic trapped 
object \cite{Friese1996}\cite{Rodriguez2009}\cite{Abbas2004}\cite{Beams1954}.

The experiment remains to be optimized, and all the data presented below are likely taken on multilayer
flakes, rather than on single layer graphene.  Nonetheless, it is hoped that further refinements 
will enable levitation of single layer graphene and that rotation
resonance measurements will effectively capitalize on the powerful techniques of analysis developed for magnetic resonance and
ultimately provide a wealth of information on the levitated sample and its internal properties.

I provide below a detailed description of the experimental design, including the development of many formulas
necessary to understand the requirements for trapping graphene, the expected optical signal from the trapped sample,
and the torques that will affect graphene motion in the trap.  After a presentation of the data, I conclude with
a consideration of possible improvements to the current design and
a discussion of applications for graphene trapping, both for fundamental measurements and as a possible new environment
for graphene crystal modification and growth.

\section{Description of the Quadrupole Trap}

At its simplest, a quadrupole trap is comprised of two or more electrodes in a configuration
where the electric field $\bm{E}$=0 at some point $\bm{r}=0$ in space away from the electrodes 
\cite{Paul1990}\cite{Dehmelt1967}.  If only 
DC potentials are applied to the electrodes then confinement of a charged particle at the point $\bm{r}=0$ is always unstable,
but stable trapping is possible with AC applied fields under appropriate conditions. The equation of
motion is:

\begin{equation}
\ddot{\bm{r}}  + \gv \dot{\bm{r}} =\frac{q}{m} \cos ( \Wtrap t) \bm{E}(\bm{r}) \mbox{,}
\end{equation}
where  $q$ is the particle charge, $m$ is its mass, $\Wtrap/2\pi$ is the frequency of the trapping field,
and dots above variables are used here and below to signify time derivatives.  
$\gv$ is the velocity damping rate, which will be extremely important for graphene flakes, due to their small mass
(typically $\sim 10^{-18}$\,kg for a $\mu$m-sized monolayer) and
large surface area to mass ratio.  

While it is possible to get exact results valid in the limit $\bm{r} \to 0$ by linearizing Eq. 1  \cite{Davis1985},
a more intuitive picture of ion trap dynamics comes from the pseudopotential
approximation \cite{Dehmelt1967}.  In this approximation the particle is assumed to be rapidly oscillating and experiences
a spatially nearly uniform electric field during each oscillation period. The position of the particle
is separated into an oscillatory ``micromotion" $\bm{\delta}$ and a position $\bm{R}$ of the particle averaged over an
oscillation period: $\bm{r}=\bm{R}+\bm{\delta}$.
Trapping occurs because in an oscillating nearly uniform electric field in the absence of
damping, $\bm{\delta}$ and $\ddot{\bm{\delta}}$ ($\propto$ force) are $180^\circ$ out of phase: a particle in the
neighborhood of $\bm{r}=0$ will experience a force \emph{towards} $\bm{r}=0$ when it is furthest away from
$\bm{r}=0$ and \emph{away} from $\bm{r}=0$ when it is closest to it.  If $\bm{E} \to 0$ at $\bm{r}=0$, then the
force on the particle averaged over its micromotion will be directed towards $\bm{r}=0$, leading to
stable trapping.  The effect of finite damping is to reduce the phase difference between $\bm{\delta}$ and $\ddot{\bm{\delta}}$,
which will tend to reduce the average trapping force directed towards $\bm{r}=0$.

The above arguments can be used to derive an equation of motion for the particle position
averaged over the micromotion \cite{Dehmelt1967}\cite{Spann2001}:

\begin{equation}
\ddot{\bm{R}}  + \gv \dot{\bm{R}} +\frac{q}{m} \bm{\nabla} \Psi (\bm{R}) =0 \mbox{,}
\end{equation}
where the pseudopotential is:

\begin{equation}
\Psi (\bm{R}) = \frac{1}{4} \frac{q}{m} \frac{1}{\Wtrap^2+\gv^2} \bm{E}^2 (\bm{R}) \mbox{.}
\end{equation}
From the formula for the pseudopotential, 
it is seen that a smaller $\Wtrap$ leads to greater confinement (provided the pseudopotential
approximation remains valid), and that the confinement rapidly weakens when $\gv > \Wtrap$. 

The trap used in the experiments described below is similar the stylus trap developed for ions \cite{Maiwald2009}.
It is a coaxial arrangement of two conically shaped
pieces of stainless steel, the apex of which is shown in Fig. 1.  The AC trap voltage is applied to the outer electrode,
while the inner electrode is kept near ground.  Additionally, the chamber surrounding the trap ($\sim$2 cm away from the apex)
is also grounded.  Calculated \cite{Femlab} values of $\bm{E}^2 (\bm{R})$ are plotted, showing the pseudopotential minimum
point about 0.8 mm away from the trap apex.  

\begin{figure}
\vspace{0cm}
\begin{center}
\includegraphics[scale=0.4,draft=false]{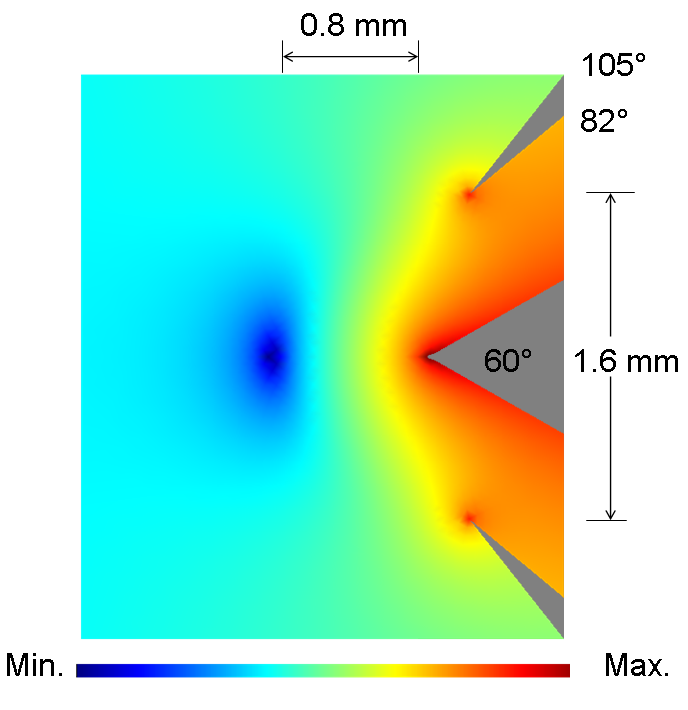} 
\end{center}
\vspace{0cm}
\caption{Cross section of the apex of the trap electrodes (grey regions) and the surrounding
pseudopotential.  The trap is made from two coaxial pieces of conically tapered stainless steel.
The oscillating trap voltage is applied to the outer electrode, while
the inner electrode is held near ground.  The surrounding vacuum enclosure (not
shown) is also grounded. The tip of the inner electrode extends beyond the face of
the outer electrode by about 200\,$\mu$m.}
\end{figure}

Along the axis of symmetry of the trap in the neighborhood of $\bm{r}=0$, $\bm{E}(\bm{r})$ may be
approximated as:

\begin{equation}
E_z= -\frac{z}{z_0^2} \Vout \mbox{~~and~~} E_x=E_y=0 \mbox{.}
\end{equation}
Here, $\Vout$ is the amplitude of the voltage applied to the outer electrode, and
$z_0$ is a parameter determined from modeling the electrode configuration,
and is 1.84 mm for the trap design depicted in Fig. 1.  When $\gv =0$, Eqs. 2-4
are readily solved to get oscillating solutions with:

\begin{equation}
\omega_z= \frac{1}{\sqrt{2}} \frac{q}{m} \frac{\Vout}{\Wtrap z_0^2}  .
\end{equation}
Similarly, for radial motion away from the axis of symmetry, oscillatory solutions can be obtained
with $\omega_x = \omega_y =\omega_z/2 $.  These solutions are obviously only valid in 
the regime $\omega_z \ll \Wtrap$.  However, the trap will remain stable when $\omega_z / \Wtrap <$0.32\cite{Pearson2006}.

Nonzero biases applied to the \emph{inner} electrode on the trap will produce an electric field at $\bm{r}=0$ of
$E_z=-\Vin /{z_1}$, where $z_1$=6.4\,mm is another parameter that can be determined by trap modeling \cite{Femlab}.
Nonzero $\Vin$ at DC shifts the trap minimum position along the $z$ axis, while $\Vin$ applied at finite
frequencies can be used to determine $\omega_z$ (and thus $q/m$) by observing resonance behavior.

Quadrupole traps are usually designed to maximize confinement while maintaining stability.  For atomic
systems, this requires that $\Wtrap > 10^7$\,\persec, and excitation at this frequency is typically provided with
a tuned resonator.  For graphene particles charged using electrospray ionization,
$q/m$=10-100\,C$\cdot$\perkg ($10^{-7} - 10^{-6}~|e|\cdot$\peramu), much smaller than the values for
atomic ions.
Consequently, the trap excitation is conveniently provided by a circuit \cite{AALab} using a high voltage op-amp
\cite{Cirrus}.  This allows for a maximum voltage amplitude of 400\,V at a maximum frequency of about 100\,kHz.
Using $\Vout$=300\,V, $\Wtrap/2\pi$=30\,kHz (values typically used in these experiments), 
and $q/m$=10\,C$\cdot$\perkg in Eq. 5 for the trap in Fig. 1,
I obtain: $\omega_z$=3300\,\persec.  Thus, the assumptions of the pseudopotential approximation are well satisfied during
the experiments discussed below.

When a particle is confined by a quadrupole trap, fluctuations about the pseudopotential minimum will ultimately be determined by
Brownian motion \cite{Marago2010}.  Using the equipartition theorem:

\begin{equation}
\frac{1}{2} \kB T = \frac{1}{2} m \omega_z^2 \langle z^2 \rangle \mbox{.}
\end{equation}
For temperature $T$=300\,K and $m=10^{-18}$\,kg, $\langle z^2 \rangle^{1/2}$= 20\,$\mu$m
for the hypothetical particle discussed above. This dimension is small compared to
the size of the trap, but it is readily measurable, and can be used to provide information
about the trapped particle.  Finally, the acceleration of gravity, $g$, will displace the particle from the trap center
by a distance $g/\omega_x^2$, which is a few $\mu$m in this experiment and is generally not
observable.

\section{Particle Damping}

From very crude kinetic theory in the free molecular (Knudsen) regime,
the velocity relaxation time of a particle, 1/$\gv$, is the time it takes to
collide with its own mass of the gas molecules in its surroundings.  For single layer graphene, this time
is just the time it would take for a monolayer of gas molecules to attach to the surface (if they all
stuck and had a mass comparable to C atoms).  The pressure-time product necessary to deposit a monolayer 
is known to surface scientists as the Langmuir unit =$10^{-6}$\,torr$\cdot$sec.  Consequently, for a graphene monolayer 
$p/\gv\cong 10^{-6}$\,torr$\cdot$sec. More refined kinetic
theory \cite{Li2003} gives a very similar result for graphene in a \N2 ambient at 300\,K.  An important
point is that for a 2D object like graphene, the mass and collision cross section are proportional to one another,
so $\gv$ is independent of the lateral size of the graphene layer.  However, The number of layers, $n$, will
affect the mass, but not cross section.  Thus, measurement of $p/\gv$ can be used to determine $n$.

\begin{figure}
\vspace{0cm}
\begin{center}
\includegraphics[scale=0.5,draft=false]{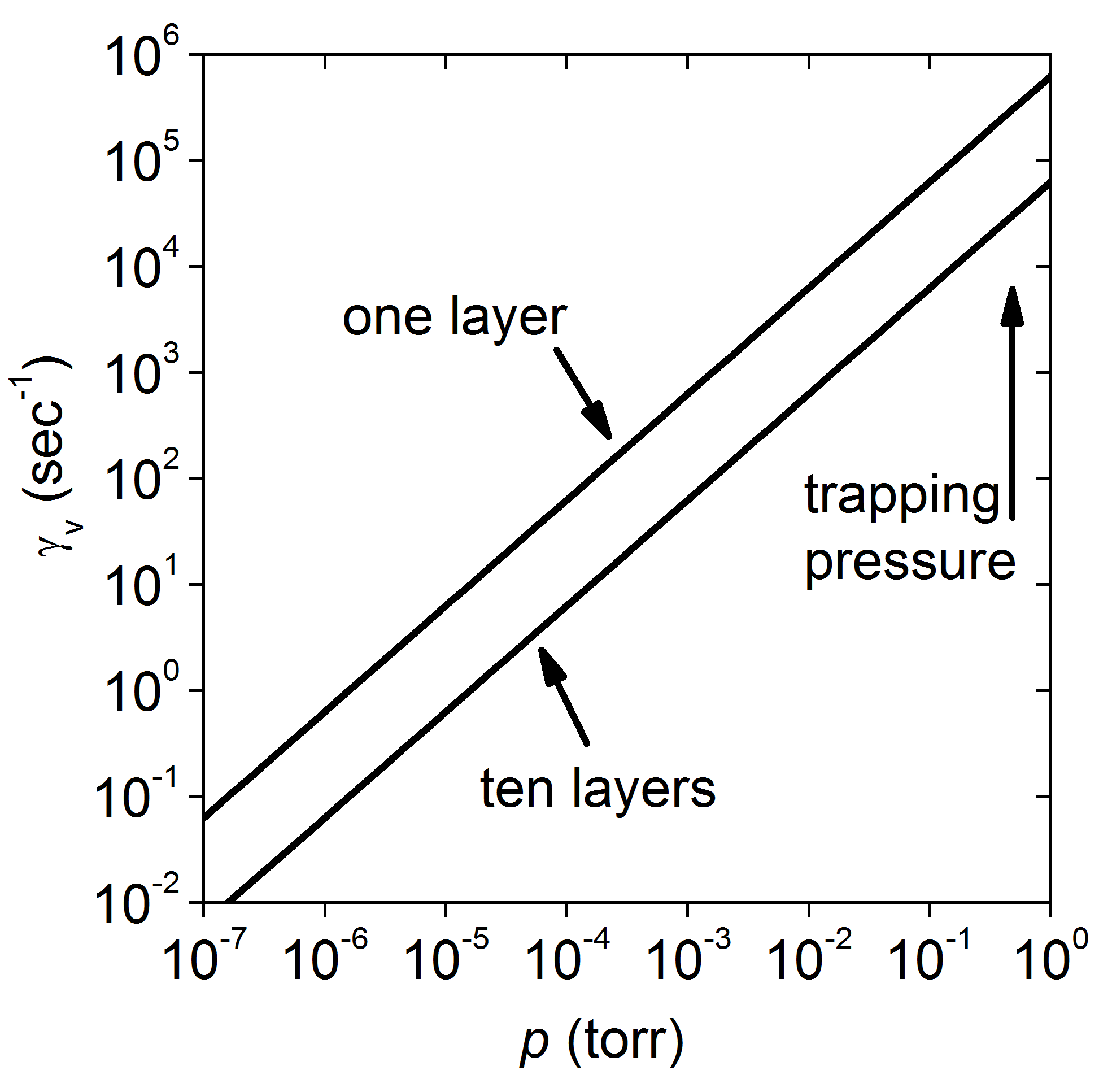} 
\end{center}
\vspace{0cm}
\caption{Predicted velocity relaxation rate, $\gv$, of graphene as a
function of pressure.}
\end{figure}

Using these ideas $\gv$ versus pressure is plotted in Fig. 2 for graphene for $n=1$ and $n=10$.  As mentioned
above, finite $\gv$ weakens the trapping pseudopotential in Eq. 3 unless $\gv < \Wtrap$.  For the trap parameter
$\Wtrap$= 200,000\,\persec, this requirement means that the trap can only operate at pressure below 1\,torr.
On the other hand, finite $\gv$ is necessary, since - without a drag force - particles could never become
bound in the trap.  Consequently, the experiments discussed below are conducted by trapping the
particles at around 0.5 torr and subsequently pumping the system down to $10^{-3}-10^{-6}$\,torr for most 
measurements.

While the picture above is adequate to motivate the design of this experiment, it is worth noting that
many of the assumptions underlying simple kinetic theory may be unwarranted for a 2D object in 
the Knudsen regime.  For spherical objects the values of $\gv$ determined assuming purely specular and
purely diffuse scattering of molecules off the surface differ by only about 15\% \cite{Li2003}.  However, for a 2D
plate  moving in a direction parallel to its surface, molecules specularly reflected 
off the surface do not transfer momentum to the plate in the direction of its motion, and thus would not contribute to
$\gv$.  Purely specular scattering would also not contribute to $\gw$ the angular velocity relaxation rate for a plate
\emph{rotating} on an axis perpendicular to the plate.  The relative contributions of diffuse and specular
scattering off graphene is unknown, but it is notable that crystalline surfaces with a high degree of perfection
are being developed as ``atom mirrors" in which the ratio of specular to diffuse scattering is significant \cite{Barredo2008}.

\section{Generation of Charged Flakes by Electrospray Ionization}

Graphene flakes are introduced into the trap using the electrospray ionization technique.  Originally developed for mass spectrometry \cite{Fenn2003}, electrospray has also been applied to inject micron-scale 
charged particles into quadrupole ion traps \cite{Cai2002},\cite{Pearson2006},\cite{Clark2009}.  In the electrospray
technique, a liquid suspension containing the particles is ejected from a capillary tube held at high voltage.  Ejected droplets
shrink in size due to a combination of liquid evaporation and droplet fission, until only dry charged
particles remain suspended in the chamber gas.  The large voltages required for electrospray mean that
it is most easily performed at or near atmospheric pressure, with charged particles subsequently 
introduced into a higher vacuum environment through a pinhole orifice located near the electrospray
emitter tip.

The suspensions used in these experiments are similar to those developed by Hernandez $et~al$\cite{Hernandez2008}\cite{Hernandez2009}.
These workers prepared suspensions by ultrasonication of graphite flakes in a variety
of liquids.  The resulting mixture was then centrifuged to deposit coarse material, while finer
particles remained suspended.  The particles that remain in suspension were shown
to be micron-scale flakes of few layer graphene, with $n$ typically in the range 1-10.

In order to use these suspensions for electrospray, it is extremely important to minimize non-volatile impurities in the 
liquid, since these impurities
will accumulate on the graphene flake during liquid evaporation, and can ultimately outweigh the
residual graphene if care is not taken.  While many liquids are effective for graphene suspensions
\cite{Hernandez2008} I have chosen to use an isopropyl alcohol (IPA)-water mixture (3:1 volume ratio),
since IPA and water are commonly used
in mass spectrometry and are available in ``MS grade" 
(low in ionic contaminants and $<$1\,ppm residue after evaporation) \cite{fishersci}.  Care must also be taken to ensure that
only clean and inert materials are brought into contact with the suspension and that it is not contaminated
with foreign particulates.
For these reasons I give a rather detailed description below of the materials
and techniques used to prepare and deliver the suspension to the emitter tip.

The suspensions are prepared in 8\,mL glass vials with PTFE-faced phenolic caps (VWR \#14230-824).
After the vials are rinsed in IPA, 5\,mg of graphite flake (Aldrich \#332461) are weighed and introduced into
the vial.  5\,mL of liquid is then added, and the cap is closed.  The vial is then placed in an
ultrasonic bath (Branson \#1510) in a position so that the fluid level in the bath matches that
inside the vial, and the sample is ultrasonicated for 30 min.  All operations where the suspension is
exposed to air are performed in a clean room.

After ultrasonication the vial is centrifuged (Drucker \#642E, approximate acceleration=1000$g$) for
30 min.  After centrifugation the supernatant has a typical optical attenuation coefficient
(measured at 532\,nm) of 80\,\perm.  Using the attenuation dependence on concentration determined by 
Hernandez\cite{Hernandez2008} of 2500 L$\cdot$\perm$\cdot$\perg, the typical concentration of graphene flakes in 
the suspension is about 30\,mg$\cdot$\perL.

\begin{figure}
\vspace{0cm}
\begin{center}
\includegraphics[scale=0.6,draft=false]{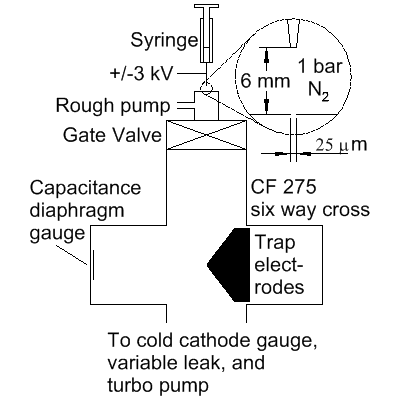} 
\end{center}
\vspace{0cm}
\caption{Diagram showing the particle injection system and the
gas handling of the experimental apparatus.  The minimum
obtainable pressure in the chamber is $\leq$1\,$\mu$torr}
\end{figure}

The supernatant is drawn through a syringe needle (Hamilton \#90122)
into a glass syringe (Hamilton \#1750, 0.5\,mL volume, with PTFE luer lock connection), which is
placed into a syringe pump (Harvard Apparatus,
Nanomite) located at the top of the quadrupole trap vertical column (Fig. 3). Upon exiting the
syringe the suspension travels through PEEK 1.6\,mm diameter tubing and fittings to the stainless
steel electrospray emitter (New Objective Corporation, 100\,$\mu$m tip inner diameter).  The metal
tip is connected to a wire that allows it to be biased at high voltages. 

During operation the emitter tip is surrounded by \N2 at atmospheric pressure and is about 6\,mm
above a 25\,$\mu$m diameter grounded stainless steel pinhole aperture (Edmund Optics). 
Because contamination may arise
from particles in the gas surrounding the emitter tip that come into contact with exiting
charged droplets, the \N2 is filtered (Swagelok, SCF Series, 0.003\,$\mu$m) prior to entering the
chamber surrounding the tip.  The chamber walls are made of glass, so electrospray emission is
visible when the tip is illuminated with a laser pointer.  Onset of electrospray emission
typically occurs at $\Vtip \cong \pm $2\,kV, with optimal performance around 3\,kV.  Optimal flow
rates during electrospray are typically 1-3\,$\mu$L$\cdot$\permin. During particle trapping a roughing
pump evacuates the chamber downstream from the pinhole to about 0.5\,torr.  After
a particle is trapped, a gate valve between the electrospray source and the main chamber is 
closed (Fig. 3), allowing the chamber to be pumped with a turbo pump to below $10^{-6}$\,torr.
For measurement of pressures $>10^{-4}$\,torr a capacitance diaphragm gauge and a convection gauge located close
to the trap center are used.  A cold cathode ionization gauge is located downstream from the trap to measure
lower pressures.  Close proximity between the trap center and the ionization gauge was avoided to prevent possible
discharging of trapped particles, and consequently there is considerable uncertainty in the pressure measurements
below $p<10^{-4}$\,torr.  Finally, an \N2 variable leak is located downstream from the ionization gauge to
enable control of the chamber pressure.

\section{Optical Apparatus}

During and subsequent to trapping, particles are imaged from light scattered at small angles from the direction
of an illuminating laser (Fig. 4).
The light source is a $\lambda=$532\,nm wavelength laser (Lasermate Corp.) with adjustable output power.
Laser power is levelled during measurements by using a photodiode and a PID controller.
Power flux at the center of the beam is determined by scanning the laser across a (non graphitic) trapped particle
and fitting the scattered intensity to a gaussian.  The power flux calculated in this way is 
3300\,W$\cdot$m$^{-2}\cdot$mW$^{-1}$.  

Care must be taken to avoid light scattering into
the collection optics that could wash out the signal.  For this reason, coated windows
(Lesker  VPZL-450AR) are used for the optical entrance and exit to the trap chamber.
Additionally, the trap outer electrode is conically tapered to minimize scatter from corners and edges.
Also, laser line filter is placed in the collection optics to minimize interference from ambient
light.

\begin{figure}
\vspace{0cm}
\begin{center}
\includegraphics[scale=0.5,draft=false]{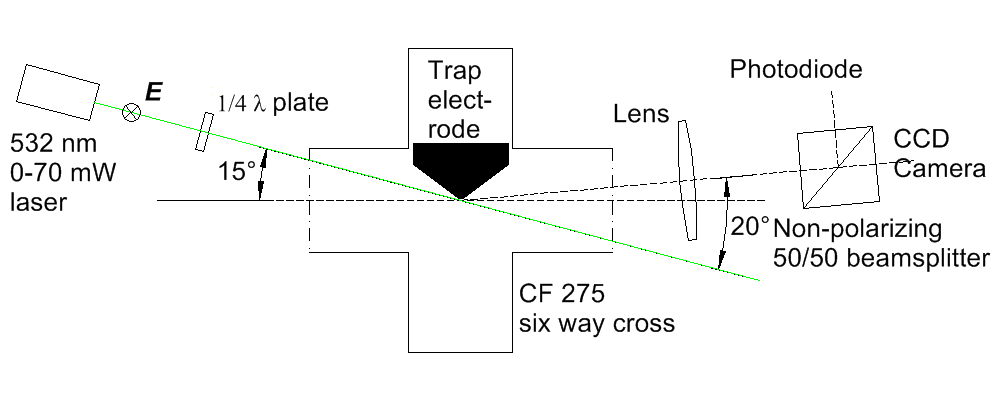} 
\end{center}
\vspace{0cm}
\caption{Diagram of optical instrumentation for the experiments.  The electric field of
the light originating from the laser points out of the diagram.  Light received at the
detectors on average has scattered $20^\circ$ from the incident beam direction. }
\end{figure}

The collection optics consists of two 40\,mm diameter lenses located outside the trap vacuum chamber,
120\,mm away from the trap center, with magnification factor of 4.2.
A beam splitter is used so that trapped particles may be simultaneously viewed by both a low
light level CCD video camera (WATEC 120N) and a low noise photodiode (FEMTO FWPR-20-SI) with a
20 Hz bandwidth.  A 500\,$\mu$m diameter
pinhole aperture is positioned in the focal plane in front of the photodiode to minimize the contribution of background
light to the detected signal coming from trapped objects.  
Finally, a $1/4$ wave plate on a rotation stage allows for control of the light polarization
illuminating the trapped particle. 

\section{Interaction of graphene with electric fields}

The graphene samples studied in this experiment are irregularly shaped flakes 
whose size will in general not be small in comparison to
the 532\,nm laser wavelength.  However, to get a rough idea of the expected signal from scattered
light, I will estimate the interactions of a graphene flake with an applied electric
field in the quasi-static dipole approximation.   In this approximation, the dipole moment,
$\bm{p}= \epsilon_0 \alpha \bm{E}$, is estimated from the polarization induced by a uniform electric field.
$\alpha$ is the polarizability tensor, and $\epsilon_0$ is the permittivity of vacuum. 
Analytically tractable results for a circular disk
may be derived by determining the polarizability of an oblate spheroid \cite{Bohren1983} 
with semimajor axis, $a$, and allowing the semi-minor
axis, $b$, to go to zero. Inside the spheroid, the complex permittivity is:

\begin{equation}
\epsilon = \epsilon_0 + \frac{\sigma_\3D}{\omega } i \mbox{,}
\end{equation} 
where $\sigma_\3D$ is the volume conductivity of the material
and $\omega$ is the angular frequency of the applied field. 
For an oblate spheroid, $\alpha_\parallel$, the response to an electric field oriented in the plane of rotational symmetry,
$E_\parallel$, is:

\begin{equation}
\alpha_\parallel = 
\frac{4}{3}\pi a^2 b \frac{ \frac{\sigma_\3D}{\epsilon_0 \omega} i }{1+ \beta \frac{\sigma_\3D}{\epsilon_0 \omega} i }  \mbox{,}
\end{equation} 
where $\beta$ is a geometical factor $\cong \pi b/(4 a)$ for $b \ll a$ \cite{Bohren1983}.
To obtain an expression for
a disk, I equate $\sigma_\2D$ with the product of $\sigma_\3D$ and $4b/3$, the mean thickness of
the spheroid in the direction of its axis of symmetry averaged over its cross section.
The result is:

\begin{equation}
\alpha_\parallel = \pi a^3 
\frac{ \left[ \frac{ \sigma_\2D }{\epsilon_0 \omega a } \right] i }{1+ \frac{3 \pi}{16} \left[\frac{ \sigma_\2D }{\epsilon_0 \omega a  }\right] i } \mbox{.}
\end{equation} 

\subsection{Optical Scattering and Absorption}

At optical frequencies it is now well established experimentally \cite{Nair2008}\cite{Mak2008}\cite{Das_Sarma2010}
that $\sigma_\2D \cong e^2/4 \hbar$ for monolayer graphene.  Using this value 
and $\omega =\wL = 3.54 \times 10^{15}$\persec (for $\lambda$=532 nm) yields a value of $4 \times 10^{-3}$ for the term in brackets
in Eq. 9 with $a$=0.5 $\mu$m.
Thus, at optical frequencies for micron-scale flakes with one or a few layers: 

\begin{equation}
 \alpha_\parallel \cong  i \pi a^2  \frac{\sigma_\2D}{\epsilon_0 \wL} \mbox{.}
\end{equation} 
This value will scale linearly with $n$, the number of layers in a flake, provided $n$ is small.
Using this approximation the optical absorption cross section, $\sigma_{abs}$, of a flake oriented perpendicular to the
incident radiation is \cite{Bohren1983}:

\begin{equation}
\sigma_{abs} = \frac{2\pi}{\lambda} \mbox{Im}[\alpha_\parallel] = n \pi a^2 \times \frac{e^2}{4 \epsilon_0 \hbar  c} 
\cong 0.023 \times n \pi a^2 \mbox{.}
\end{equation} 
This picture predicts that a graphene monolayer absorbs 2.3\% of the light impinging on its area, in agreement with
experiments \cite{Nair2008}. This absorption has significant implications for optical measurements, since sample heating
can be large in a high vacuum environment at laser powers above a milliwatt (Fig. 5).  

\begin{figure}
\vspace{0cm}
\begin{center}
\includegraphics[scale=0.5,draft=false]{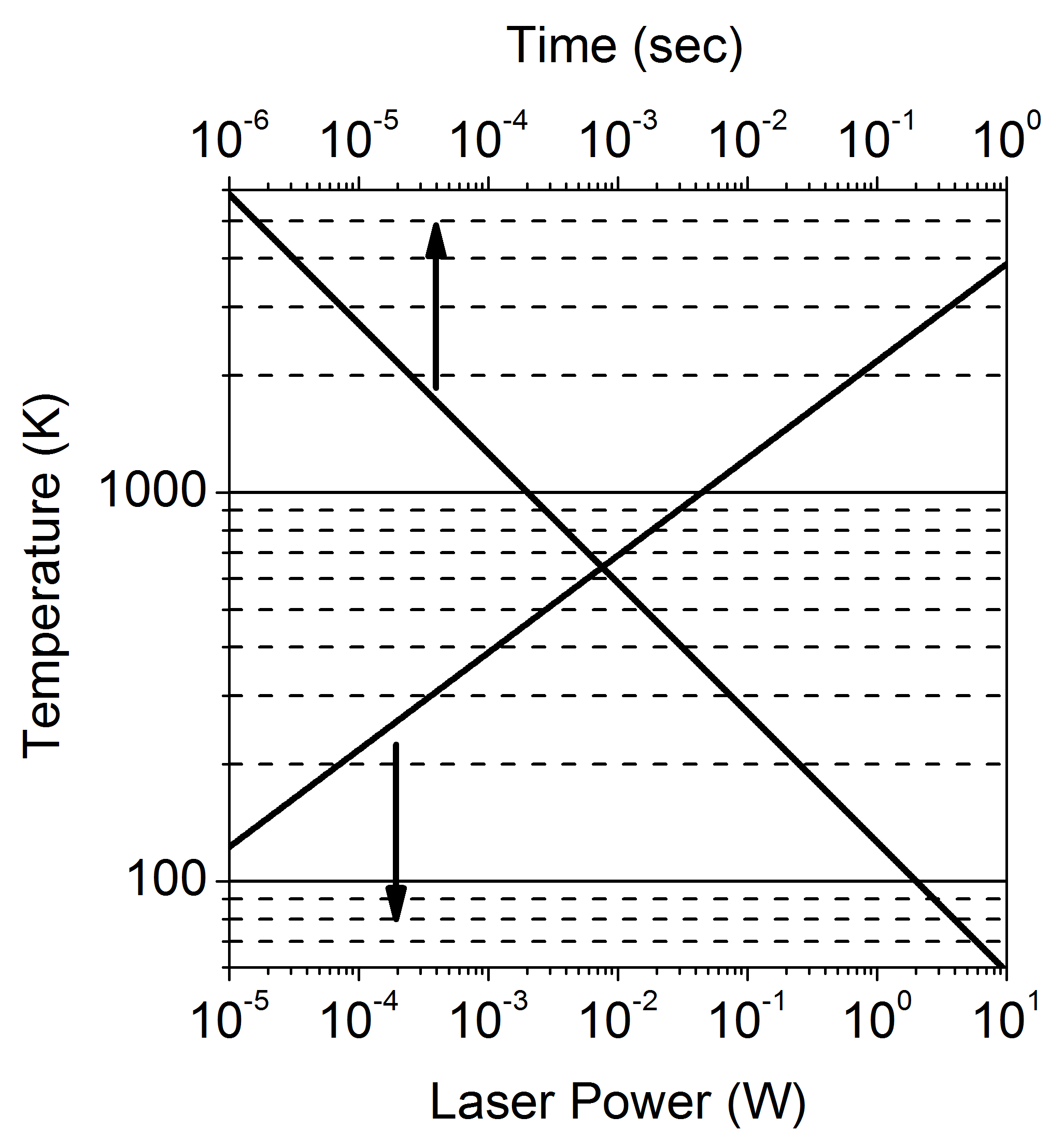} 
\end{center}
\vspace{0cm}
\caption{Predicted heating (versus laser power) and cooling (versus time after the heating source is turned off) 
of a graphene single layer in vacuum.  Heat is assumed to be removed from the flake by
black body radiation into a surrounding environment at $T=0$, and the heat capacity is assumed to be 3$\kB$ per C atom.  
The laser beam diameter is 0.5\,mm, and 2.3\% absorptivity and emissivity are assumed. Size effects and temperature
dependence of the heat capacity have been neglected and will become increasingly important at low temperature.}
\end{figure}

To determine the visibility of a flake I next calculate the forward scattering differential (per solid angle $\Omega$)
cross section of a flake oriented
perpendicular to the incoming radiation using the same assumptions \cite{Bohren1983}:

\begin{equation}
\dcs [\theta=0] = \left(\frac{1}{4 \pi}\right)^2 \left( \frac{2\pi}{\lambda} \right)^4 |\alpha_\parallel|^2 
\cong  \left(\frac{0.023 \times  n \pi a^2}{2 \lambda} \right)^2  \mbox{.}
\end{equation}
The quantity $n \pi a^2$ is proportional to the mass of the flake.  The result of Eq. 12 is plotted in Fig. 6 as a function of particle
mass using typical experimental parameters.  These calculations show that a single layer of graphene
with a 1\,$\mu $m$\times$1\,$ \mu $m area should be visible in the experiments.  It is important to note, however, that visible objects
will not conform to the condition that $a \ll \lambda $.  This, combined with the fact that flakes will be irregular in shape
(and possibly contain regions with different layer thickness), means that these results are at best rough estimates of the
expected optical scattering from trapped flakes.

\begin{figure}
\vspace{0cm}
\begin{center}
\includegraphics[scale=0.5,draft=false]{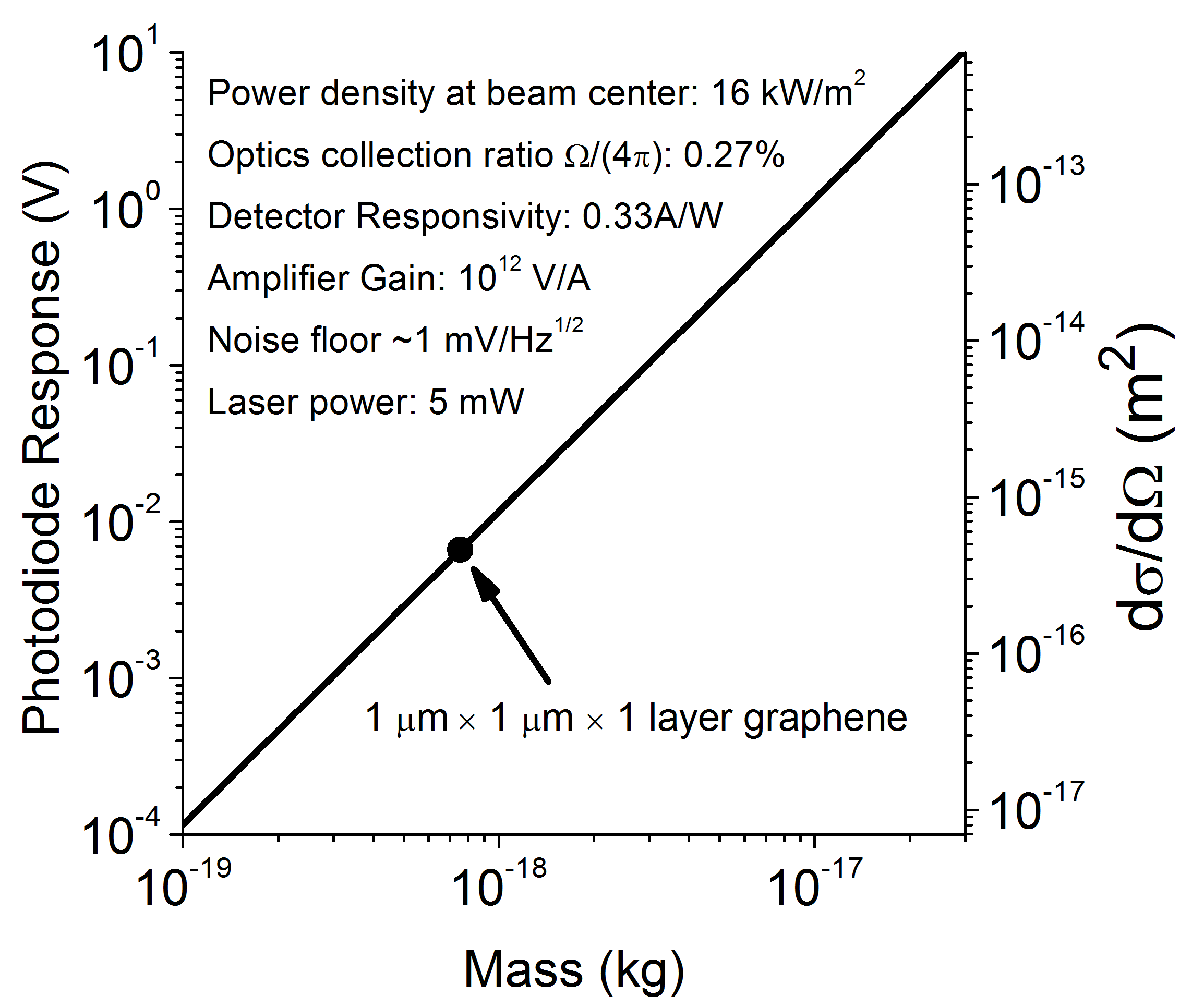} 
\end{center}
\vspace{0cm}
\caption{Predicted optical signal  and forward scattering cross section for graphene as a function of its size.
Finite size corrections to dipole scattering are neglected. }
\end{figure}

\subsection{Optical Anisotropy}

The results above were all derived for a flake in a plane oriented
perpendicular to the direction of the incoming radiation, when $\bm{E}$ always lies in the plane of the flake.
Because of the thinness of graphene (and the anisotropy of graphite in general), 
$\alpha_\parallel  \gg \alpha_\perp \approx 0$.  Consequently the optical
response of trapped graphene should vary with time if the flake orientation changes with respect to the
incident radiation.  If the flake is randomly oriented in linear polarized light (LPL):

\begin{equation}
\dcs^{\mathrm{mean}}_{\mathrm{LPL}}= \frac{2}{3} \dcs^{\mathrm{max}} \mbox{~~~and~~~}
\dcs^{\mathrm{min}}_{\mathrm{LPL}}= 0 \mbox{.}
\end{equation}
The maximum value is that determined from Eq. 12, and will be the same for linear or circular polarized incident
light.  Because circular polarized light (CPL) effectively averages over orientation around the direction of incident
radiation:
\begin{equation}
\dcs^{\mathrm{mean}}_{\mathrm{CPL}}= \frac{2}{3} \dcs^{\mathrm{max}} \mbox{~~~and~~~}
\dcs^{\mathrm{min}}_{\mathrm{CPL}}= \frac{1}{2} \dcs^{\mathrm{max}} \mbox{.}
\end{equation}
These equations predict that temporal fluctuations should approach 100\% in LPL and 50\% in CPL.  The
slow (20\,Hz) response of the photodiode in the experiments means that these fluctuations will be averaged
out at high pressures, however.

\subsection{Optical Torque}

Because electromagnetic waves carry angular momentum as well as energy, absorption of light (Eq. 11) must also
convey torque, $\bm{N}=\bm{p}\times\bm{E}$, to the particle if the incoming radiation is circularly polarized:

\begin{equation}
N_{\perp} =  I_\perp \dot{\omega}_\perp = \pm \frac{S}{\wL}\times 0.023 \times n \pi a^2  \mbox{,}
\end{equation}
where $S$ is the incoming radiation power flux.  $N_\perp$, $\omega_\perp$, and $I_\perp$  are respectively the torque, angular velocity,
and moment of
inertia associated with motion around an axis perpendicular to the plane of the flake, which is assumed to be oriented
perpendicular to the direction of incident radiation.
For a circular flake with uniform density: 
$I_\perp = m a^2/2= \pi n a^4 \rho_\2D /2$, where $\rho_\2D$ is the 2D mass density of single layer graphene=7.6 
$\times 10^{-7}$\,kg$ \cdot \mathrm{m}^{-2}$.
Thus:
\begin{equation}
\dot{\omega}_\perp \cong \pm \frac{S}{\wL} \times 0.023 \times \frac{2}{a^2 \rho_\2D}  \mbox{.}
\end{equation}
Note that since both torque and mass are linearly dependent on $n$, $n$ does not appear in this formula.
For 5\,mW laser power and $a\cong$0.5\,$\mu$m, $\dot{\omega}_\perp=1.1 \times 10^6$\,$\mbox{sec}^{-2}$.
In circularly polarized light, the angular rotation velocity of a graphene flake will increase until
the optical torque is matched by frictional drag at $\omega_{\mathrm{max}}= \dot{\omega}_\perp/\gw$
This result implies that at low pressures, when 
$\gw \le $1\,\persec, rotational frequencies above 1\,MHz are possible.  The \emph{thermal}
rotational angular velocity is given by:

\begin{equation}
\frac{1}{2} \kB T = \frac{1}{2} I \omega^2 \mbox{,}
\end{equation}
which, for a 1\,$\mu$m diameter single layer graphene flake at 300\,K, leads to $\omega =2 \times 10^5$\,\persec.
Thus, at low pressures, light-induced rotation will exceed thermal rotation of the graphene flakes.

\subsection{Low Frequency Torques}

At frequencies $\le$1\,GHz, the bracketed term in Eq. 9 $\gg$1 and:

\begin{equation}
\alpha_\parallel \cong \frac{16}{3}  a^3 
\left( 1+ \frac{16}{3\pi} \left[ \frac{\epsilon_0 \omega a }{ \sigma_\2D } \right] i \right)  \mbox{.}
\end{equation} 
The real part of this expression is simply the static electric polarizability of a thin disk of
radius $a$ \cite{Friedberg1993}. This term produces a torque along an axis in the plane of the flake
when $\bm{E}$  is at an angle $\theta$ from the flake plane:
\begin{equation}
N_\parallel = E_{\perp} \epsilon_0 \alpha_\parallel E_\parallel  
 = \frac{16}{3} \epsilon_0  a^3 E^2 \sin \theta \cos \theta = I_{\parallel} \dot{\omega}_\parallel \mbox{.}
\end{equation}
For 2D objects the
moments of inertia have the property: $I_{\parallel 1}+I_{\parallel 2}=I_{\perp}$, and for a circular uniform disk:
$I_{\parallel}=m a^2/4$.  Thus, for a disk:
\begin{equation}
\dot{\omega}_\parallel = \frac{64}{3} \frac{a}{m} \epsilon_{0} E^2 \sin \theta \cos \theta \mbox{.}
\end{equation}
Typical low frequency electric fields that can be conveniently applied in the trap (by applying a voltage
to the \emph{inner} electrode) are $\sim$ 1000\,V$\cdot$\perm.  This value is comparable to the electric field that
the particle experiences due to thermal motion away from the trap minimum.  For $a$=0.5\,$\mu$m and $m=10^{-18}$\,kg, 
$\dot{\omega}_\parallel$ is at most $5 \times 10^7 \mbox{sec}^{-2}$, a value that can be considerably larger than the value for
the angular acceleration from optical fields calculated above.

Finally, the imaginary term in Eq. 18 leads to a braking effect, since dissipation
occurs when charge moves on the spinning flake to shield external electric fields oriented in the flake plane.
For a circular disk:
\begin{equation}
\gw(E_\parallel)=\dot{\omega}_\perp/\omega_\perp=\frac{512}{9\pi} \frac{a^2}{m \sigma_\2D } \epsilon^2_0 E^2_\parallel \mbox{.}
\end{equation}
At low frequencies, it is inappropriate to use the optical value for $\sigma_\2D$, since the conductivity
will depend on the number of carriers in the flake (and possibly on quantum size effects in a mesoscopic
system).  However, to obtain an order of magnitude estimate of the
braking rate, I use $\sigma_\2D = e^2/4 \hbar$, $a=$0.5\,$\mu$m, $m=10^{-18}$\,kg, and $E$=1000\,V$\cdot$\perm to get 
$\gw(E_\parallel)\cong 10^{-5}$\,\persec.  From a perusal of Fig. 2 it is seen that $\gw(E_\parallel)$ is negligible compared
to damping from background gas (assuming $\gv \cong \gw$) for the conditions of this experiment.
However, damping either from applied
electric fields or those experienced during thermal motion could become dominant at pressures $<10^{-10}$\,torr.

\section{Measurements}

While the graphene suspensions described above are stable for weeks, for these experiments they are
typically prepared within 24 hours of their use.  After the syringe containing the suspension is placed on the
trap column, high voltage is applied to the tip, and the gas surrounding it is purged with filtered \N2 for several
minutes.  The gate valve is then opened to the trap chamber and liquid flow is initiated out of the electrospray
tip by activating the syringe pump.  Once electrospray emission is initiated, particles are observable with the
video camera (typically 0.1-1\,\persec), and pressure in the trap chamber increases.  The  500\,mtorr pressure in the
chamber used during trapping will decrease, along with collection efficiency, as the pinhole becomes contaminated
during operation.

\begin{figure}
\vspace{0cm}
\begin{center}
\includegraphics[scale=0.2,draft=false]{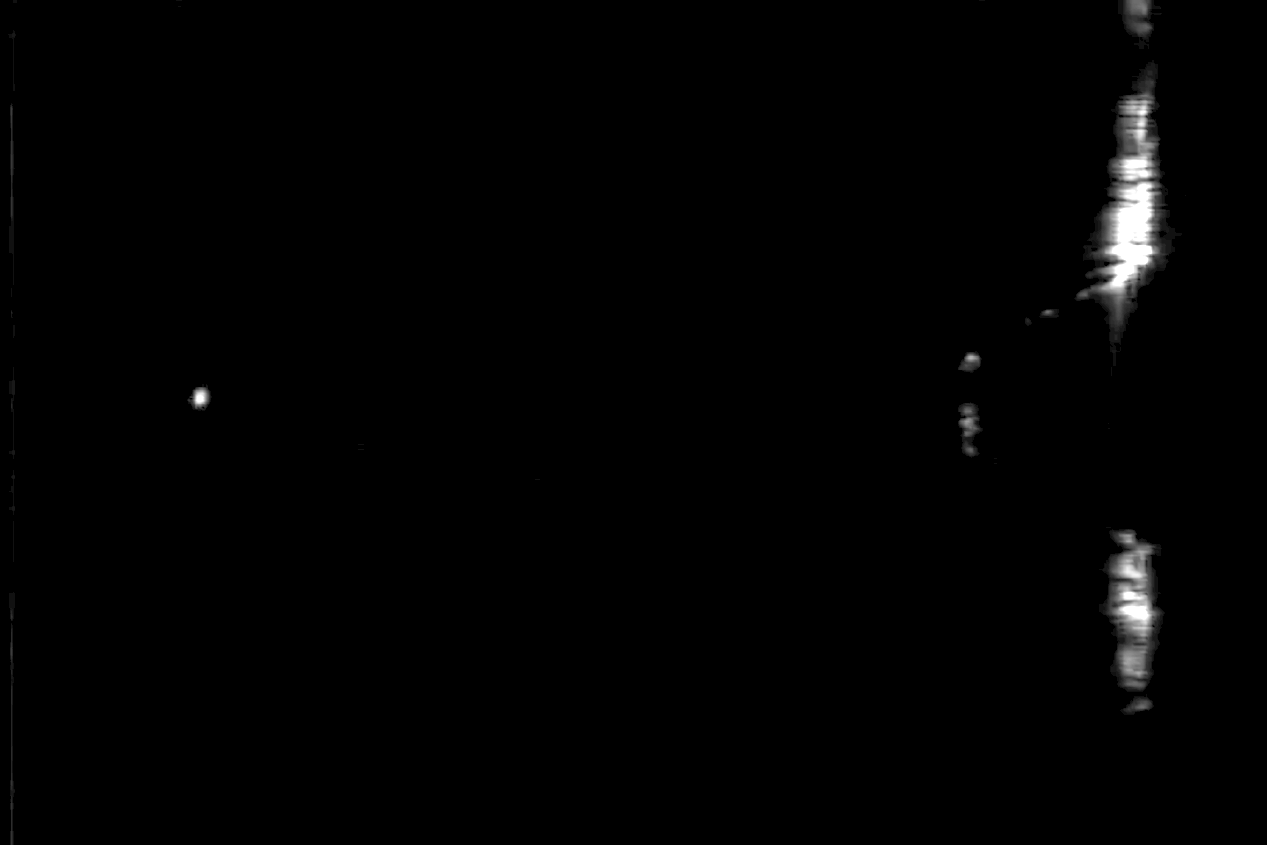} 
\end{center}
\vspace{0cm}
\caption{Photo of the trap with a confined visible graphene flake.
The particle is on the left, while scattering from the faces of the
electrodes is visible on the right. }
\end{figure}

Trapping of particles passing near the trap is facilitated by applying a DC bias (typically $\pm$2\,V) to the inner electrode
that draws the trap minimum towards the electrodes.  Once a particle is caught, the gate valve is closed; the DC
bias can be removed; and the chamber can be pumped to lower pressures with the turbo pump.  Trapped graphene
flakes are highly sensitive to light: at high pressures (when they are trapped) this can be seen in ``fluttering" 
(irregular motion) that increases with laser power; at lower pressures ($\le$ 10\,mtorr) particles can rapidly discharge
(and leave the trap) when the laser power significantly exceeds 10\,mW.  These effects are consistent with heating of the flakes
by absorbed light.  Discharging at high light levels occurs at a similar rate for both negatively and positively charged
trapped flakes, suggesting that thermionic emission is not the operative mechanism.  Most likely ionic species are present
on the flake surface of both polarities and volatilize at high temperatures to discharge the flake.

At the lowest obtainable pressures ($p<$0.1-1\,$\mu$torr) damping is extremely inefficient, and particles 
are frequently expelled from
the trap, most likely from electrical or acoustic noise coupling to the resonant frequency of motion of
the trapped flake.  At higher pressures and at low light levels,
particles can remain in the trap for weeks. 
 
During pumping to low pressures, the optical signal coming from scattering from the flake
is monitored with the photodiode.  Strong fluctuations in time appear typically at $p<$10\,mtorr
for graphene flake samples (Fig. 8a).  These fluctuations are largest in magnitude when the
sample is viewed with LPL.  Such fluctuations do not appear when non-graphitic
dust particles are trapped.  Of the particles that do display fluctuations, greater than half
of them display virtually 100\% signal modulation in LPL at $p<0.1$\,mTorr.
Only these ($\sim 100\%$ modulating) are chosen for more detailed study.  It should be emphasized at the
outset that this criterion does not mean that the graphene flakes are single layer, since even
multilayer flakes are still optically very thin, and should display a large anisotropy. It is
unlikely, however, that crumpled balls or wads of graphene would display large anisotropy, and
consequently the measurements discussed below are taken on thin, flat flakes. 

\begin{figure}
\vspace{0cm}
\begin{center}
\includegraphics[scale=0.7,draft=false]{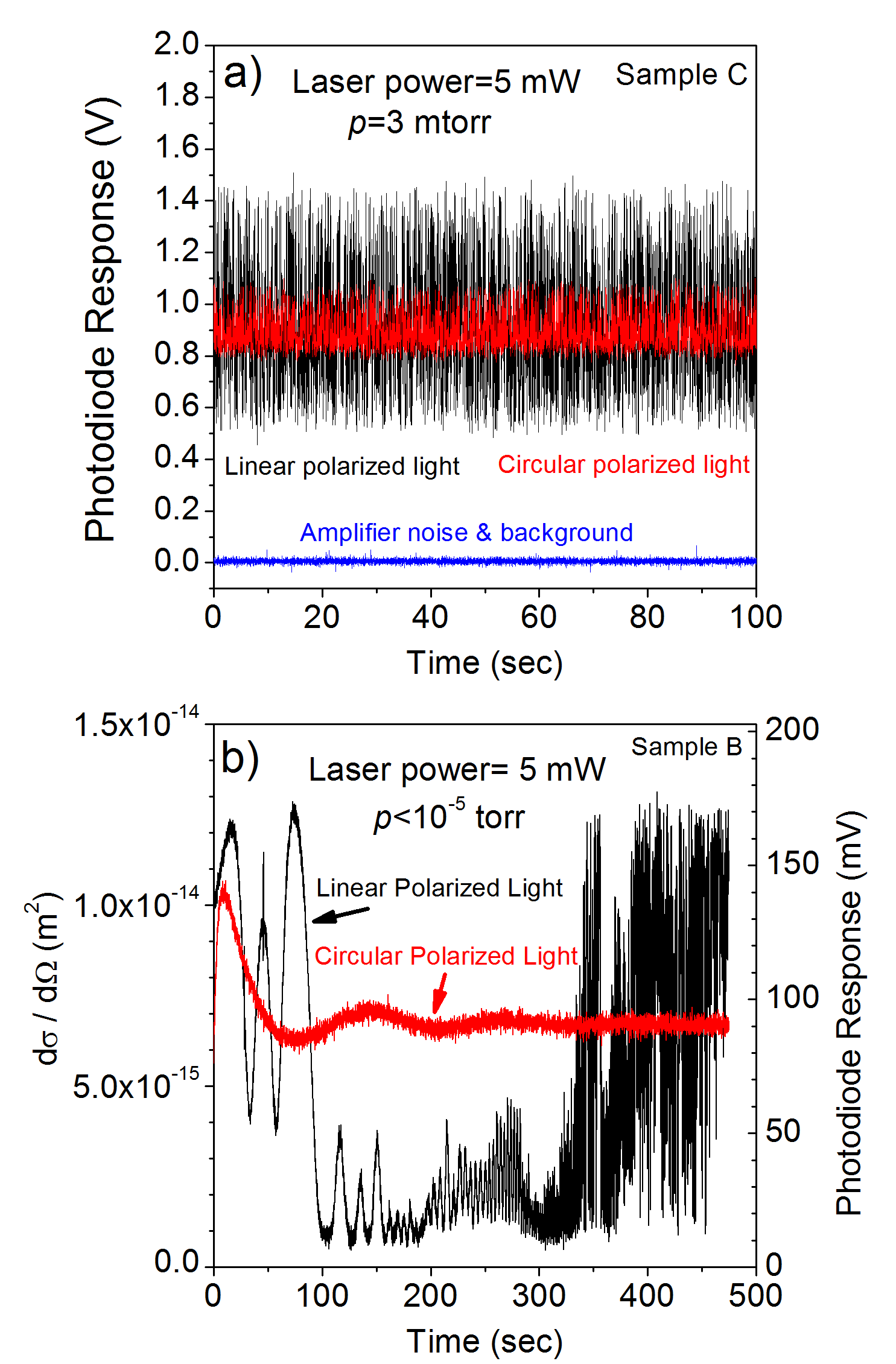} 
\end{center}
\vspace{0cm}
\caption{(a) Light scattering from a graphene flake at a pressure near the onset of the appearance of fluctuations.
At higher pressure the scattered light signal becomes uniform.  
(b) Behavior of a flake at low pressure.  For the circular polarized light
measurement, the laser is turned on at $t=0$ after being off for an extended period.
Data is taken for linear polarized light after switching away from a long exposure to
circular polarized light at $t=0$.  The minimum signal observed using linear polarized light
is limited by instrumental noise.}
\end{figure}

Data for a such a flake at $p<10^{-5}$\,torr is shown in Fig. 8b.  When pumped down while viewing in
LPL, the fluctuations evident at higher pressures increase to 100\% of the
photodiode signal.  If the sample is then illuminated with CPL, the fluctuations
almost completely disappear, and the signal approaches a nearly constant, equilibrated value.  If
the illumination subsequently is switched back to LPL, quasiperiodic behavior is observed: the characteristic frequency of the 
fluctuations increases with time until they exceed the measurement bandwidth of the photodiode,
at which time large amplitude random fluctuations are again observed.

The likely interpretation of this behavior is that rotation of the flake is induced by CPL,
which stabilizes its orientation relative to the direction of incident radiation.  The quasiperiodic
behavior observed when the illumination is switched to LPL
is a consequence of the gradual slowing of this rotation due to friction
with the residual gas in the trap chamber.  Periodic behavior can occur if there are torques on the flake
along an axis in the flake plane, which can come from electric fields, as was discussed above.  These torques
would cause the axis of rotation of the flake to precess with angular velocity $\sim \dot{\omega}_\parallel/\wg(t)$.
The precession frequency increases as the angular velocity of the graphene flake, $\wg$,
slows down, just as the wobbling frequency of a top increases as
it spins down.

It has been shown under general circumstances \cite{Xu2008} that the axis of symmetry of small, 
absorbing, oblate spheroid will tend to align with the direction of incident radiation
when illuminated with CPL.  The plane of the flake is thus presumably
oriented perpendicular to the incident radiation when the particle has reached its steady
state value when illuminated with CPL.  The theory above predicts that the scattered
light intensity should be maximal in this orientation, since the electric field is always in the plane
of the flake.  The data in Fig. 8b clearly show, however, that peak scattering occurs under LPL illumination and
substantially exceeds the scattering from the flake after it has been illuminated by CPL for a long period.  
This discrepancy is likely an indication that the flake is
not in fact significantly smaller than a wavelength.  A flake in the mirror configuration, where the angle of incidence
is equal to the angle of scattering and where the
electric field lies in the flake plane, should presumably scatter with peak intensity for a large flake.
It is likely that this is the orientation of the flake when the signal is maximal under LPL illumination.

The characteristics of five samples that showed rotating behavior at low pressures are listed in
Table 1.  $q/m$ is determined by observing motional resonance (typically at 1-10\,mtorr) and using Eq. 5 with
the known trap parameters.  $\dcs^{\mathrm{max}}_{\mathrm{LPL}}$ is the peak signal
scattered from the sample in LPL, and  $\dcs^{\mathrm{equil}}_{\mathrm{CPL}}$
is the signal from the sample after it has been illuminated with CPL for an extended period.
Both of these optical measurements are made at $p<10^{-5}$\,$\mu$torr.
Because it is likely to be least sensitive to size effects, $\dcs^{\mathrm{max}}_{\mathrm{LPL}}$
is used to determine the ``optical" mass using Eq. 12, and $q$ is then determined from the optical mass.  

\begin{table}
\begin{center}
\begin{tabular}{|l|l||c|c|c|c||c|} \hline
Parameter&Units&A&B&C&D&E\\
\hline
$\Vtip$&kV&-2&-2.2&2.8&2.8&3.1\\
$\Wtrap/2\pi$&kHz&35&30&20&30&20\\
$q/m$&C$\cdot$\perkg&-121&-48&16&30&14 \\[0.5 ex]
$\dcs^{\mathrm{max}}_{\mathrm{LPL}}$&$10^{-15}$\,m$^2$&
5.5&12&160&100&460\\[0.5 ex]
$\dcs^{\mathrm{equil}}_{\mathrm{CPL}}$&$10^{-15}$\,m$^2$&
3.3&6.5&75&50&210\\[0.5 ex]
Optical $m$&$10^{-18}$\,kg&2.6&3.9&14&11&24\\
Inferred $q$&$|e|$&-1970&-1170&1400&2060&2100\\
\hline 
$\gw$&\persec&0.02&0.012&0.003&0.01&0.09\\
$p$&$\mu$torr&0.6&1.0&0.3&0.6&1.5\\
$p/\gw$&$\mu$torr$\cdot$sec&30&80&100&60&17\\
\hline
\end{tabular}
\end{center}
\caption{Characteristics of five flake samples.    $\Vtip$ is the voltage applied
during electrospray ionization and determines the sign of the charge of the trapped particle.
Optical mass is that inferred from Eq. 12 and Fig. 6.  The rotation resonance measurements were
performed on Sample E.}
\end{table}

It is also possible to estimate the mass by observing Brownian motion (also typically at 1-10\,mtorr) and using Eq. 6.
This technique gives results 3-10 times smaller than the optical mass.  Systematic errors can easily enter
this measurement, however:  $T$ for the flake under illumination is not known accurately.  Also, electronic
and acoustic noise can contribute to the observed fluctuations and cause the mass to be underestimated.

Perhaps the most surprising aspect of the data is the extremely long times it takes for the sample to
equilibrate at low pressures.  The rotational damping rates $\gw$ listed in Table 1 are estimated from
the changing frequency of the quasiperiodic oscillations observed during spin down in LPL,
assuming that the torque that is causing precession is constant. The pressure reading on the ionization
gauge downstream from the trap was between 0.3 and 1.5\,$\mu$torr during these measurements. The resultant values
for $p/\gw$ are large and suggest that the samples all contain many layers if the simple kinetic picture of damping
presented above is correct and if $\gw \sim \gv$.  The pressure measurement in the experiment is certainly
inadequate, but to bring the data into line with theory for a monolayer flake, the pressure in the chamber would need to be
\emph{lower} than the gauge value, which would be surprising.  (The cold cathode gauge reading is closely matched
to the capacitance diaphragm gauge, very near to the trap center, when they are both in range.) 

\section{Observation of Rotational Resonance at MHz Frequencies}

The data presented above provides strong indirect evidence that flake spinning is induced by exposure to
CPL. The long measured values of $\gw$ - combined with the estimates of
optical torque for graphene - suggest that the rotation frequency is in the MHz range in the experiments.
Direct optical observation of this spinning using fast optical detectors is a challenge due to the small (femtowatt) signals
scattering off the flakes.  An alternative is to excite the flakes at high frequencies and seek to observe
changes in the optical signal measured at low frequencies.  In particular, a sample can be exposed
to high frequency electric fields by applying a voltage to the \emph{inner} electrode of the trap.
At MHz frequencies, these fields have a negligible effect on the position of the trapped flake,
but do produce torques on the sample as was discussed above.

\begin{figure}
\vspace{0cm}
\begin{center}
\includegraphics[scale=1.0,draft=false]{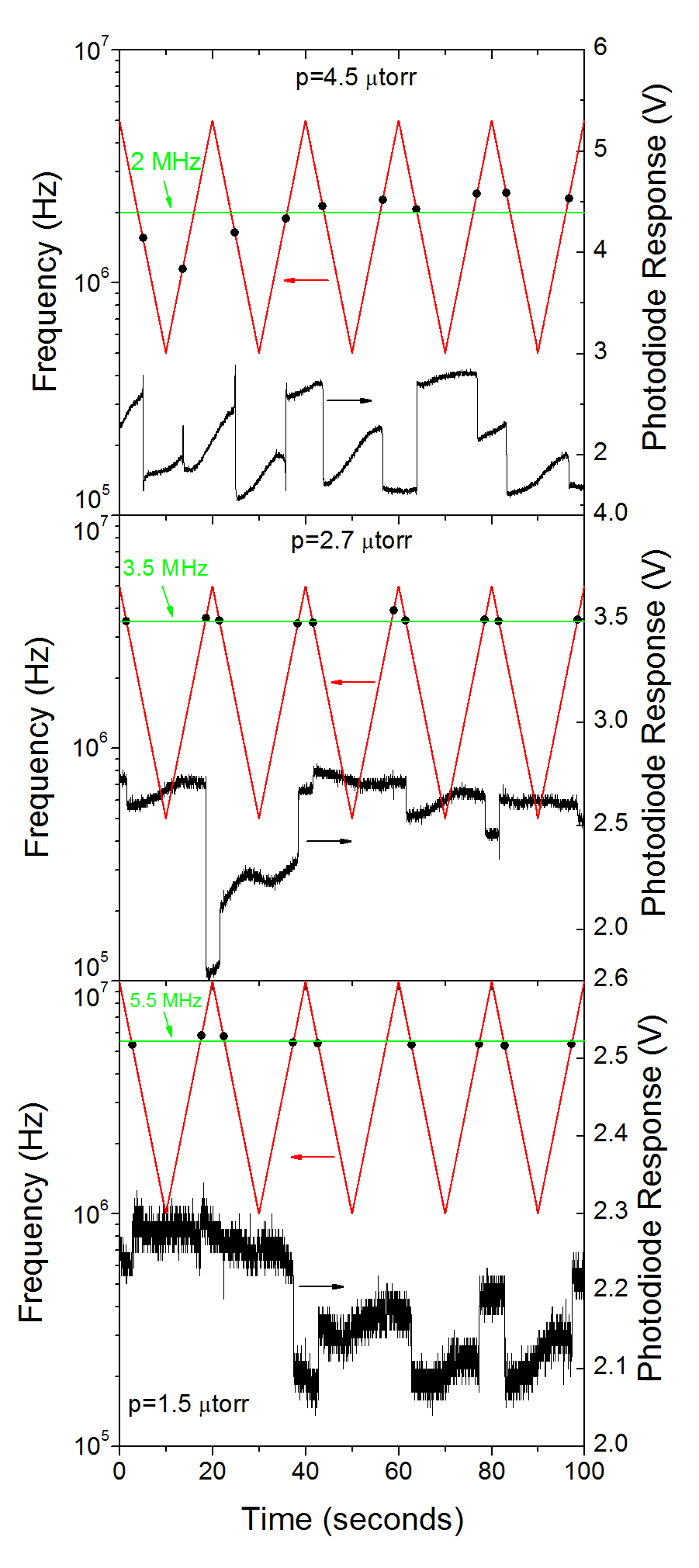} 
\end{center}
\vspace{0cm}
\caption{Observation of rotational resonance of a graphene flake.
For all of the data, the sample is exposed to 4\,mW circular polarized light during
the measurements.}
\end{figure}

To perform this experiment a sample was illuminated with CPL at $p<10$\,$\mu$torr.
An 8\,V amplitude AC signal was applied to the inner electrode, corresponding to a peak $E$ at the
trap center of 1250\,V$\cdot$\perm. The frequency was ramped logarithmically while the optical scattering signal
was monitored (Fig. 9).  The data clearly show sharp jumps in the photodiode signal, occurring both when the frequency is
scanned upwards and downwards. The jumps are not uniform in size, however.  Note also that the
direction of the jump (towards brightening or dimming) is not always the same for a given direction of
the frequency sweep.

\begin{figure}
\vspace{0cm}
\begin{center}
\includegraphics[scale=0.7,draft=false]{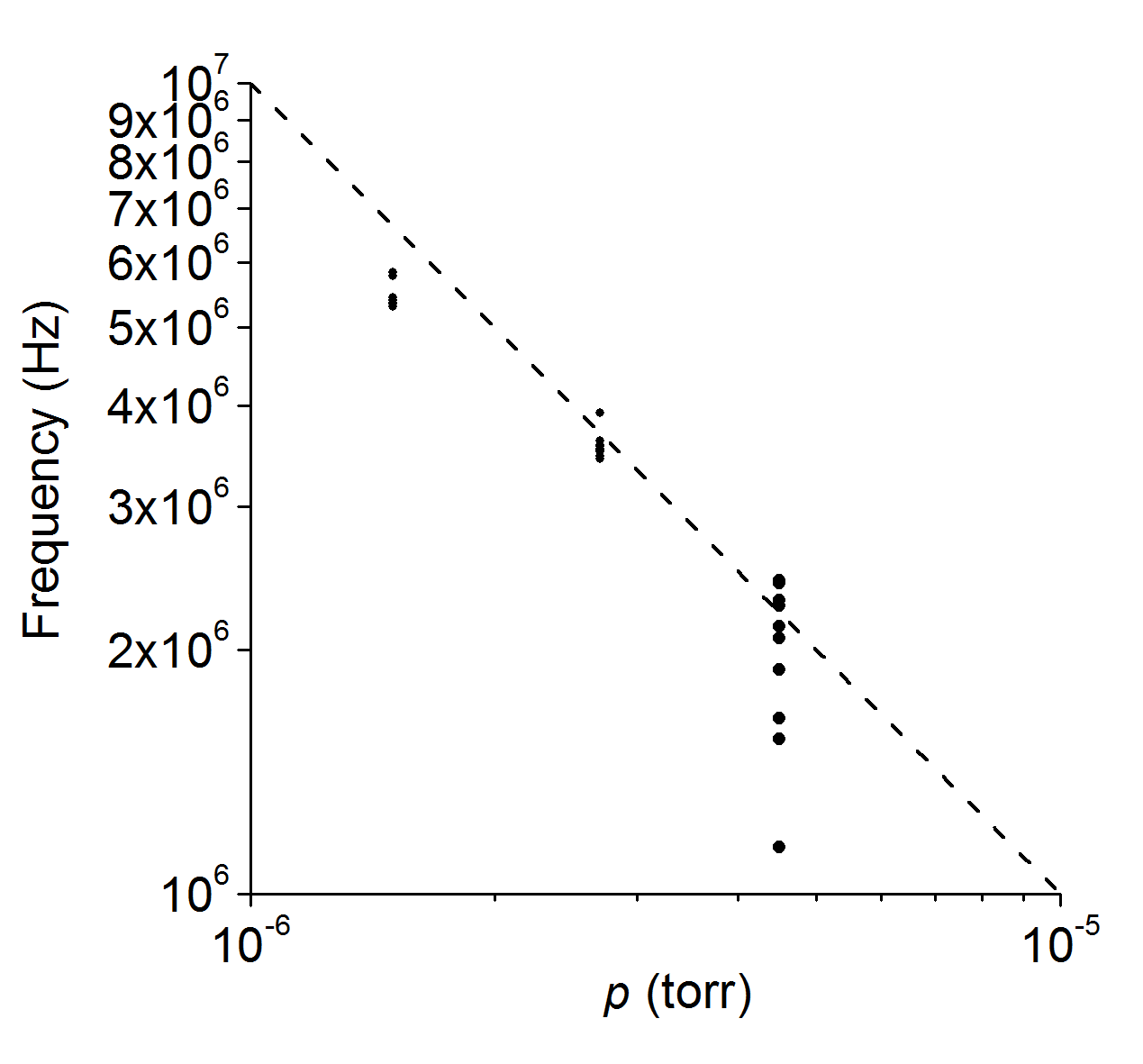} 
\end{center}
\vspace{0cm}
\caption{Effect of chamber pressure on the frequency at which rotation resonance jumps
occur.}
\end{figure}

The rotation resonance jumps show clear dependence on the pressure in the trap:  the magnitude of the jumps
decreases markedly at low pressure, and the frequency at which the jumps occur is roughly inversely proportional
to the pressure (Fig. 10).  The scatter in the frequency associated with the jumps is increasing at high
pressures, probably an indication that random or thermal fluctuations in the rotation and orientation of
the flake are becoming significant.  Subsequent to these resonance measurements, the bias applied to
the inner electrode was switched off, and $\gw$
was determined using observation of the spin down behavior in linear polarized light (Fig. 11).

\begin{figure}
\vspace{0cm}
\begin{center}
\includegraphics[scale=0.7,draft=false]{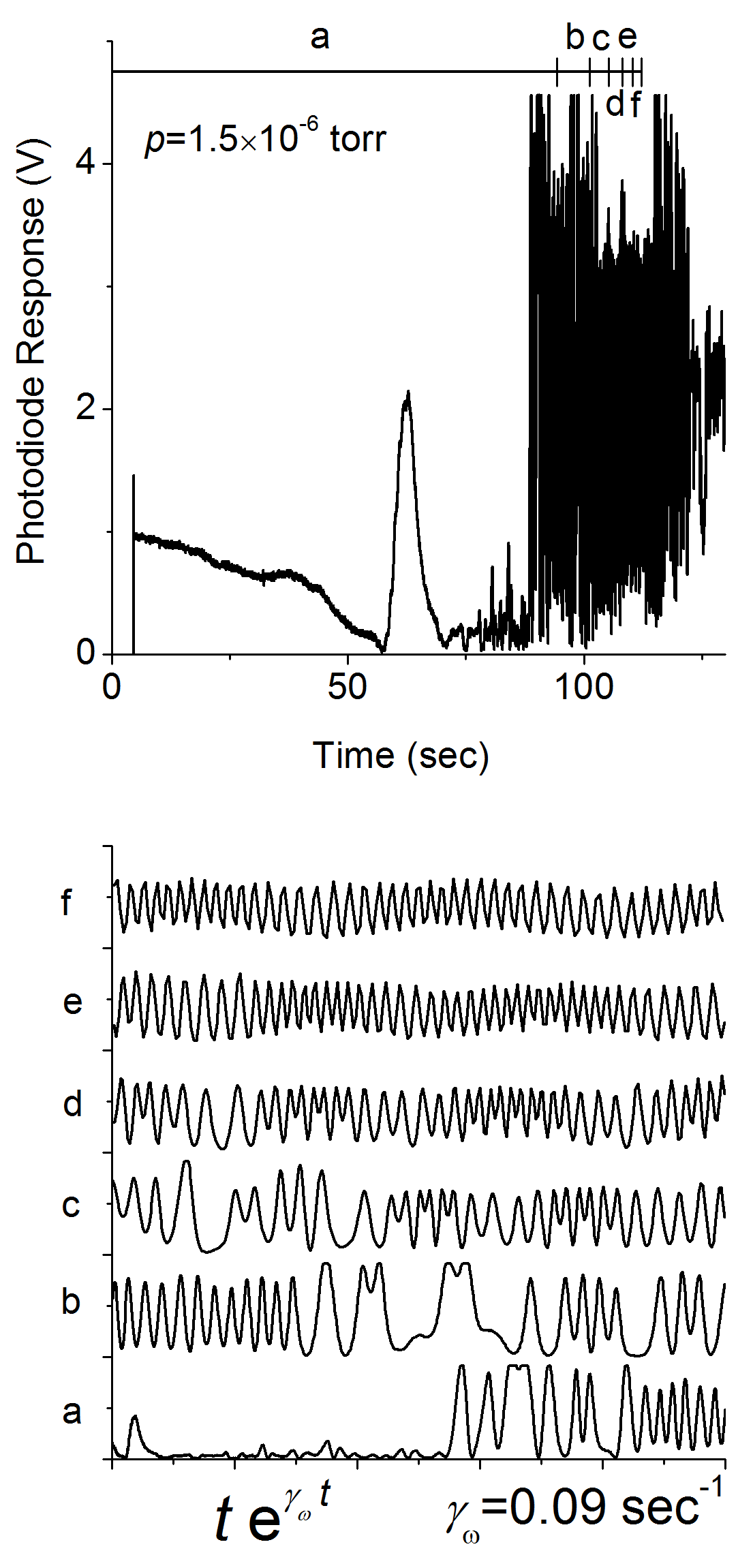} 
\end{center}
\vspace{0cm}
\caption{Spin down behavior measured subsequent to observation
of rotational resonance of the flake observed with  4\,mW of linear polarized light.  
Previous to $t=0$, the flake had been exposed to 4\,mW of circular polarized light
for a long period.  To facilitate the observation of the quasiperiodic oscillations,
the time axis is exponentially expanded in the bottom plot.}
\end{figure}

The fact that signal intensity changes abruptly strongly
suggests that the axis of orientation of the flake (or its average orientation) is
shifted by a resonant interaction with the applied torque.  While
it is tempting to treat this as a problem of rigid body dynamics, a better approximation for graphene 
is to examine the dynamics of
a thin membrane with no rigidity whatsoever under tension purely from centrifugal forces.  This problem 
was treated by Lamb and Southwell \cite{Lamb1921} for the case of a circular disk.  The lowest frequency
transverse excitations of the disk spinning with angular velocity $\omega$
have a single nodal diameter that rotates at $\pm \omega$ in the frame moving with the
spinning disk.  In 
a nonrotating frame, these excitations correspond to a DC tilt of the axis of
rotation and fast precession at $2\omega$ of the rotation axis of the disk.
This latter motion can presumably be excited by an external torque applied at
angular frequency $2\omega$.  

For \emph{rigid} 2D plates of arbitrary shape (where $I_{\parallel 1}+I_{\parallel 2}=I_{\perp}$) , 
small deviations of the instantaneous axis of rotation from the principal axis
perpendicular to the plate precess at angular velocity \cite{Marion}: 
\begin{equation}
\omega \sqrt{\frac{(I_{\perp}-I_{\parallel 1})(I_{\perp}-I_{\parallel 2})}{I_{\parallel 1} I_{\parallel 2}} }=\pm \omega
\end{equation}
in the co-rotating frame of reference, just as was the case for the spinning circular membrane.
It is likely then that under very general conditions, rotating 2D objects are susceptible to
transverse excitation at angular frequency $2\omega$.

Equating the observed resonance frequency in the data with twice the rotation frequency of the flake,
$\wg/\pi$, allows for the estimation of the flake dimensions:  $\gw$ may be estimated from the data in Fig. 11.  Combining these values
(both measured at 1.5\,$\mu$torr) allows for the determination of 
$\dot{\omega}_\perp=\wg\gw=1.7\times 10^6$\,$\mbox{sec}^{-2}$.
Eq. 16 can then be used to evaluate the size of the flake: $a\cong$0.4\,$\mu$m. 
Furthermore, $p/\gw$ can be used to estimate $n$=17, if
the simple kinetic theory of damping is valid.  A flake with these
dimensions would have $m=\pi a^2 n \rho_\2D=6\times 10^{-18}$\,kg,
a factor of four lower than the optical mass listed in Table 1.

Finally, using these estimates for the particle size and the result of Eq. 20, the precession rate of the axis
of the flake induced by torques from the resonant electric field can be estimated: 
$\dot{\omega}_\parallel/\wg \sim$0.3\,\persec (at $p$=1.5\,$\mu$torr).  
Resonant reorientation (on the slow experimental time scales)
is thus a reasonable explanation for the observed modulation of light scattering from the rotating graphene flake
at specific excitation frequencies.  A detailed understanding of the reorientation dynamics observed in
Fig. 9, however, will require additional experimentation.

\section{Optimizing the Measurements}

The measurements presented above are obviously preliminary, and many aspects of the experimental design need
to be improved.  More uniform  samples  and better techniques to estimate their dimensions are
desirable.  A more accurate model of light scattering from micron-scale graphene flakes will also be
required.  Below is a list of several improvements to the experiment.

\subsection{Trap Design}

The trap design presented in Fig. 1 suffers from the disadvantage that there is a large cubic term in
the confining pseudopotential near the trap minimum.  When excited to large amplitudes the system
exhibits hysteresis, making it difficult to measure the frequency of resonance accurately.
Traps with a symmetric design can make much more precise determinations of 
$q/m$ \cite{Cai2002} and (by observing single electron discharging or charging events) of $q$  and $m$ 
separately \cite{Schlemmer2004}.  An advantage of the current design, however, is that a trapped
particle can in principle be transferred between two traps whose apexes are brought close together.
Sample transfer may be useful, for example, in a load lock to a UHV or cryogenic environment. 

\subsection{Particle Injection and Preparation}

While the techniques presented above for creating graphene suspensions and
injecting them into the trap have yielded encouraging preliminary results,
the data suggest that that flakes trapped so far have been multi- (or even many)
layer.  It is possible that the suspension contains mostly multilayer flakes.
However, it is also likely that there is substantial selection bias in the experiment
towards trapping large particles:  given a $\sim$30\,mg$\cdot$\perL suspension of graphene
injected into the trap chamber at 1 $\mu$L$\cdot$\permin, there are $10^5-10^6$ particles
injected every second, while typically $\sim$0.1-1\,\persec are observed to enter
the trap vicinity.  If the trap design favors trapping heavy particles (for example
if the chamber pressure is so high that the large damping of light particles prevents
effective trapping), then measurements of trapped particles will not reflect the
distribution of sizes of particles in the original suspension.  Further measurements
on both the distribution of particle sizes in the suspension and of biases present in
the trapping design are necessary.

In the current experiment $q/m$ is not a readily adjustable parameter (aside from its
sign).  Further charging of trapped particles would be desirable, for example, to offset
the discharging of the particles that was observed at high light intensities or to
increase flake charge beyond what is possible using the electrospray ionization method.  Additional charging could be
provided, either by using ultraviolet light \cite{Abbas2006} or by an electron beam \cite{Schlemmer2004}.

Finally, while suspensions of graphene created using ultrasonication yield irregularly shaped flakes
with a distribution of sizes, substantial improvement are possible: with a starting material of long cylinders etched using
deep reactive ion etching (RIE) \cite{Zhang2005} from highly
oriented pyrolytic graphite (HOPG), for example, the lateral dimensions of
the flake could be made uniform.  Ultra-centrifugation techniques \cite{Green2009} may lead to the production of suspensions
containing flakes of only a specified number of layers.

\subsection{Optics}

The optical setup presented above will clearly benefit from a more rapid and versatile method for
controlling light polarization.  The measurements of rotational resonance were performed
under continuous CPL illumination for simplicity.  A better experiment would illuminate
with CPL to confer angular momentum on the flake and measure with LPL in order to maximize sensitivity to flake orientation.
Another improvement would
be to view scattering at an angle perpendicular to the incoming radiation (and thus in the plane of the
spinning disk) to maximize sensitivity of scattering to small changes in the orientation of the flake.
  
\section{Applications for Levitated Graphene}

\subsection{Mechanical and Materials Properties Measurements}

Graphene has been deemed the world's strongest material because of its large Young's modulus and proven
ability to withstand tensile strain in excess of 10$\%$ \cite{Lee2008}.  Spinning graphene up to high rotational
velocities should provide an important measurement tool for graphene in an environment where uniform and controllable
tensile stresses can be applied.
The tensile stress from centrifugal force, $f_c$, at the center of a circular graphene single layer
rotating about an axis perpendicular to its plane
is \cite{Lamb1921}: 
\begin{equation}
f_c=\frac{1}{8}(3+\nu)  a^2 \omega^2 \rho_\2D \mbox{,}
\end{equation}
where $\nu$ is the Poisson ratio (around  $\sim$0.17 \cite{Lee2008} for graphene).
If the yield strength is roughly at 10$\%$ strain and the Young's modulus is $E_\2D$=340\,N$\cdot$\perm \cite{Lee2008}, 
then the maximum possible value of $a\omega\sim10^4$\,m$\cdot$\persec,
and the maximum rotation frequency of a graphene flake with $a$=1\,$\mu$m is $\sim$1.7\,GHz.
While this rotation rate is much greater than was achieved in the experiments presented
above, higher rotational velocities should be attainable, either by performing experiments at lower
pressures or at higher laser powers.

In addition to centrifugal force, charged graphene also experiences electrostatic tension, $f_e$, 
which may be estimated from the capacitance ($C$) of a circular disc \cite{Friedberg1993}:

\begin{equation}
f_e=\frac{d}{d(\pi a^2)} \left(\frac{q^2}{2C} \right) = \frac{1}{2 \pi a} \frac{d}{da} \left(\frac{q^2}{16 \epsilon_0 a} \right) =
\frac{q^2}{32 \pi \epsilon_0 a^3} \mbox{.}
\end{equation}
For the experiments discussed above, $q \sim  \pm 2000$\,e and $a \sim$0.5\,$\mu$m so $f_e \sim 10^{-3}$\,N$\cdot$\perm.
While this number suggests that flake charge can be substantially increased, it is likely that electronic or
ionic field emission at the edges of the flake will determine the maximum charge it can hold, rather than
the intrinsic strength of the graphene.

The measurement of rotational resonance, in addition to providing direct information about the flake's
rotation frequency, may provide a means to measure dissipation of vibrational excitations of the spinning disk
when coupling to the external environment is extremely small and controllable. Rotational
resonance may have sensitivity exceeding mechanical resonance of graphene attached to substrates
for determination of the mechanical properties of the material, especially in the regime of large tensile deformations.

The thermal isolation of trapped graphene, combined with its strong optical absorption, means that
it should be relatively easy to measure graphene's material and chemical properties at extremely high temperatures
without any interaction from a substrate.  Very little is known about the melting of graphene \cite{Geim2009}
or how defects will behave at high temperatures.  Temperature dependent adsorption and desorption of various
atomic species \cite{Ishii2008} introduced into the trap chamber can presumably be measured with high
accuracy, either by measurement of the flake mass \cite{Schlemmer2004} or by rotational resonance.
 
\subsection{Crystal Growth and Modification of Trapped Graphene}

A significant limitation of the approach to graphene trapping presented above is that flake size is
limited: it is improbable that flakes larger than $\sim$10\,$\mu$m can be injected into the trap using
liquid suspensions and the electrospray technique.  Trapped flakes may possibly be modified \emph{in situ},
however.  For example,
high temperature anneals of rotating flakes might alter the shape of the trapped flake by promoting migration and smoothing
rough edges.  It is possible that the trapped environment may also have applications for
graphene crystal growth: the general arguments against 2D crystallization \cite{Mermin1968} 
are unlikely to be valid in the presence of electrostatic
or centrifugal tensions that would tend to keep the structure planar.
The low mass of graphene monolayers means that even ``wafer-scale" graphene can
be held in traps without difficulty.  The challenge is to find appropriate conditions where a small trapped
graphene crystal will expand and maintain crystallinity when exposed to carbon sources, such as those used for C doping
in molecular beam epitaxy \cite{Manfra2005}\cite{Schmult2005}.  One possibility is that vacancies are first injected
into the bulk from the edge by a brief exposure to a high temperature.  Subsequently, at a lower temperature,
C from an impinging molecular beam is incorporated into the crystal at the positions of the vacancies.  Because
of the short time constants associated with cooling graphene at high temperatures (Fig. 5),
such multi-temperature growth cycles could proceed very rapidly.

The fact that electrostatic or centrifugal tensioning of graphene flakes can be used to increase the lattice
constant by up to 10$\%$ may also facilitate novel heteroepitaxial materials based on a
graphene substrate with an adjustable lattice.  Boron nitride, with a lattice constant about 2$\%$ greater than graphene,
may grow conformally on an appropriately tensioned substrate.  Possible C-BN multilayers grown in this
way may modify the band structure of the carriers in the graphene \cite{Giovannetti2007} or
improve their mobility \cite{Dean2010}.

Is it likely that for many applications a sample prepared or modified in a trap will ultimately need
to be positioned on a substrate.  It is possible that electromagnetic focusing techniques (like those
used in mass spectroscopy and electron microscopy) could be used with trapped flakes to place them 
accurately on substrates, although maintaining proper orientation of the flake during its deposition onto a surface could
be challenging.

\subsection{New Physics}

The strong optical absorption that makes observing and spinning graphene easy will unfortunately make if
very challenging to cool samples and observe new physics, either of the mechanical system or of the 
confined charge carriers.
The optical measurement scheme presented above would need to be
dramatically improved to enable low temperature measurements.  Nonetheless it is worth mentioning 
some experimental possibilities worthy of investigation: buffer gas cooling of the trapped flake using
$^3$He may enable cooling to $\sim$0.3\,K.  Measurement of the 2D charge carriers could
be accomplished either by their effect on the rotational resonance behavior or by the braking effect noted
above (Eq. 21).  Graphene samples annealed at high temperature to remove adsorbates and measured in high vacuum
may have extremely high carrier mobilities \cite{Das_Sarma2010}.  Traps could be
placed in magnetic fields to facilitate measurements in the quantized Hall regime, but the charge
density on the flakes will not be uniform  \cite{Friedberg1993}
and will be sensitive to the orientation and the rotational
velocity of the flake.  Torques associated with orbital diamagnetism \cite{Koshino2007} of the 2D charge carriers should be
detectable by their effect on flake orientation \cite{Zhu2003}.  It is also possible that under appropriate 
conditions resonant coupling between electron or nuclear
spins situated in the flake and the rotation or orientation of the flake could be observable.

Finally, it may be possible to cool levitated graphene using unorthodox techniques: for example,
adiabatic detensioning of the vibrational modes of
a spinning membrane may lead to cooling as the rotational frequency is slowed.
Low temperatures could conceivably be reached either by coupling to a laser-cooled atomic
system \cite{Treutlein2007} \cite{Zipkes2010} or by direct optical cooling using 
cavity optomechanics techniques \cite{Thompson2008} \cite{Chang2010} \cite{Singh2010}.
Perhaps the ultimate goal is to reach $kT/\hbar<\wg$, a regime where thermal excitations of the rotating
membrane are suppressed, analogous to cooling to the ground state of a mechanical system \cite{O'Connell2010}.  
For a graphene flake rotation frequency of order 1\,GHz, this goal can be reached at 
$T\sim$ 50\,mK.  Trapped graphene would then be a ``spinning qubit" for studies in quantum
information science.

\section{Acknowledgements}

This work was supported by the Laboratory for Physical Sciences.
The author has benefited from discussions with M. Fuhrer and C. Monroe.
Special thanks to B. Palmer for the use of a table in his lab.

\end{document}